\documentstyle[12pt]{article}
\begin{document}
\begin{center}
\baselineskip 0.70cm
\vskip 0.4in
{\Large \bf Topological asymmetry in the damping-pairing contribution
of electron-boson scattering} 
\vskip 0.3in
{\large \bf
G. Varelogiannis}


{\it Institute of Electronic Structure and Laser \\
Foundation for Research and Technology - Hellas\\
P.O. Box 1527, Heraklion, Crete 71110, Greece}
\vskip 0.15in

and
\vskip 0.15in

{\large\bf M. Peter}

{\it
DPMC,                                                  
Universit\'e de Gen\`eve\\ 
24 quai E. Ansermet,
CH-1211 Gen\`eve 4, Switzerland}

\vskip 0.6in
\begin{abstract}
\baselineskip 0.80cm

We make a
harmonic analysis of the pairing and damping contribution
of a finite $k$ range isotropic electron-phonon (or other boson) scattering in 
an anisotropic two-dimensional electronic system. We show
that the pairing contribution of the anisotropic part of
the electronic system can be much larger than its damping contribution
enhancing significantly $T_c$.
The higher is the order of the harmonic of the electronic anisotropy,
the higher can be the asymmetry in its damping-pairing contribution.
This could explain the puzzle of
a much broader quasiparticle peak in the n-doped 
than in the p-doped cuprates, their smaller $T_c$'s being also attributed to larger
damping effects. 
\end{abstract}
\vskip 0.9cm
\vskip 0.5cm
PACS numbers: 74.25.Jb 74.20.-z \hfill
\end{center}
\newpage
\baselineskip 0.90cm

The two-dimensional character of the electronic properties of 
cuprates is reflected in the structure of their Fermi surface.
Measurements by Angular Resolved Photoemission Spectroscopy 
(ARPES) \cite{Reports,Onellion} and Positron anhilation \cite{Peter}
indicate that the Fermi surface corresponding to the $CuO$ planes 
is almost cylindrical. This is also the result of band structure 
calculations \cite{Pickett}.
On the other hand some phase sensitive experiments in high-$T_c$
superconductors report the possibility of the variation of sign of the
order parameter in the $ab$ plane \cite{Phase}. There is intense theoretical 
activity on the study of the high-$T_c$ phenomenology
in terms of d-wave pairing that could originate for example 
from the interaction
of electrons with antiferromagnetic
spin fluctuations \cite{AF}. 

An alternative interpretation of the
anisotropies in terms of finite range in $k$ space
electron-phonon scattering has been proposed 
recently by several groups in different contexts 
\cite{Anderson,Abrikosov,RC2,Ruvalds,pr9507052,pr9511139,Weger}.
Such momentum modulation of the electron-phonon scattering
could result from the strong Coulomb correlations 
of the carriers \cite{Kulic} reflecting the possibility of
a phase separation instability \cite{crete} in Hubbard type models,
or could be simply due to the two dimensional character of the electronic system
\cite{Pickett2}. We will consider here an analytically solvable model
for an isotropic finite range (in k-space) electron-phonon
scattering 
in a two dimensional electronic system which respects tetragonal
symmetry.
Our analysis is not specifically linked to phonons and is valid
if phonons are replaced by other bosons provided the requirements
of adiabaticity are fulfiled.
Making an harmonic analysis of this model, we obtain an unexpected
interference between electronic anisotropies and superconducting
and damping properties. We show that the anisotropic part of
the electronic system can contribute much more to the pairing
than to the damping effects and this of course enhances significantly $T_c$.
We also obtain a simple understanding of characteristic differences 
between the n-doped and p-doped materials.
Our analysis points out that within conventional theory of superconductivity,
{\it sharp variations of the electronic density of states can
strongly favor high-$T_c$}.

We consider a two-dimensional electronic system  
respecting tetragonal symmetry in the $ab$ plane. 
We also consider that the characteristic momenta exchanged during the pairing 
interaction are small compared to the characteristic momenta of the 
variations of the electronic Density of States (DOS) over the
Brillouin zone. In that case we have different couplings in 
different regions of the Fermi surface, which are proportional to the local
DOS \cite{pr9507052}. Anisotropies in superconductivity 
reflect the DOS anisotropies since the electron-phonon 
matrix element is supposed for 
the moment isotropic.
The momentum dependent Eliashberg function \cite{reviews} can be written
in the form of Fourier series respecting tetragonal symmetry
\cite{Allen76}
$$
\alpha^2 F
\approx \sum_{M=0}^{\infty} F_{4M}(\Omega)\cos (4M\phi')
\exp\{-A|\phi-\phi'|\}
\eqno(1)
$$
The characteristic range of the interaction is $\phi_0=1/A$.
This type of Eliashberg function corresponds to the models 
considered in References \cite{Abrikosov,pr9507052,pr9511139,Weger},
details being in fact irrelevant.
We consider here
an exponential sharp cut-off for the momentum range of the
interaction in order to obtain analytic results,
however the form of the cut-off is not expected to have relevant implications
as it has already been checked numerically in Ref. \cite{pr9511139}.
The $M=0$ term represents the homogeneous or isotropic part of the DOS.
This term is dominant when the van Hove singularity is far from the Fermi
level and the DOS is rather isotropic. When for example by doping the van Hove
singularity is 
pushed close to the Fermi level at an energy distance of the order $\Omega$,
the anisotropic terms of the interaction ($M\neq 0$) become relevant. 

In an analogous way the superconducting gap respects tetragonal symmetry 
and can also be written as a sum of Fourier series
$
\Delta=\sum_{M'=0}^{\infty}\Delta_{\bar{M}}cos(\bar{M}\phi)
$
where $\bar{M}=4M'$ in the case of anisotropic s-wave gap
or $\bar{M}=4M'+2$ in the case of an anisotropic d-wave gap.
Both types of gap are accessible with an electron phonon coupling
as that considered in eq. (1) depending in fact on the exact value and 
momentum structure of the Coulomb pseudopotential $\mu^*$.
It is for example possible to have transitions between the two types of gap
by adjusting the doping since $\mu^*$ is very sensitive on it \cite{pr9511139}.
With these notations it is not difficult to see that the renormalization
due to the electron-phonon interaction becomes proportional to the 
integral
$$
I_Z(\phi)=\int_0^{2\pi} d\phi'
{\sum_{M=0}^{\infty}F_{4M}\cos\bigl(4M(\phi-\phi')\bigr)
\exp\{-A|\phi-\phi'|\}
\over
\sqrt{\omega_m^2+\bigl[\sum_{M'=0}^{\infty}\Delta_{\bar{M}}
\cos(\bar{M}\phi')\bigr]^2}}
\eqno(2)
$$

There is a non trivial relationship between the effective anisotropies of
the interaction, and its pairing and damping contributions that
has never been exploited up to now.
We consider first the damping effects in the normal state.
The general belief is that the larger is the interaction,
larger are the damping effects no matter the topology of the electronic
system. Here we will study for the first time the interference of the
topology of the electronic system with the damping effects.
In the normal state,
the damping integral of equation (2)
gives the following simple result
$$
I_Z(\phi)=
\sum_{M=0}^{\infty} F_{4M} {2A\over A^2+(4M)^2} \cos(4M\phi)
\eqno(3)
$$
The coefficients $F_{4M}$ are specific parameters 
of a given electronic system, while the intrinsic properties
of the anisotropic part of the interaction  
are contained in the ratio $2A/(A^2+(4M)^2)$.
We can make a very important remark here.
The lower is the order of the DOS harmonic (the lower is $M$) the
larger is the amplitude of its contribution
to the damping because of the presence of $(4M)^2$ in the
denominator. This effect is more 
pronounced when $A$ is smaller (when the range of the interaction is larger).
Therefore the damping contribution of the electron-phonon 
scattering is strongly dependent on the topology of the electronic 
system. High DOS anisotropy harmonics are 
irrelevant for the normal state damping effects
or in other terms high DOS harmonics can couple with any boson field
(including phonons) without affecting the effective
mass of the carriers !

If for example by doping we brink the van Hove singularity closer to the
Fermi surface enhancing $F_{4M\neq0}$, then contrary to the
general belief {\it we do not enhance} 
in a significant
way the damping effects in the normal state because of this destructive
interference of the anisotropies of the electronic system
with damping.
We show in figure 1 the contribution to the damping effects
of the $M=0$, $M=1$ and $M=2$ harmonics as a function of 
the characteristic range of the interaction $\phi_0$. Our approach is
relevant to the order $1-\sin(\phi)/\phi$ and is therefore 
reasonable for angles
up to at least $20^o$. We can see that for $\phi_0>5^o$, the larger $M$
harmonics begin to give significantly smaller damping contributions.

Having interactions that do not contribute to the damping is a very     
positive situation for superconductivity provided that these interactions
contribute to the pairing.
In fact, taking into account in a first approximation strong coupling effects, 
we can write 
$
T_c\propto \exp\{(1+\lambda_Z)/\lambda_{\Delta}\}
$
where
$\lambda_Z$ is the damping contribution of the Eliashberg function
and $\lambda_{\Delta}$ its pairing contribution. In isotropic 
superconductors we have $\lambda_{\Delta}\approx\lambda_Z$. If now the 
high harmonics do not contribute to the damping but give a significant
contribution to the pairing we may obtain 
$\lambda_Z\ll \lambda_{\Delta}$ which is the optimal condition
for high-$T_c$. 

Near $T_c$
the pairing contribution $\lambda_{\Delta}$ can be shown to be proportional
to the integral
$$
I_{\Delta}(\phi)=\int_0^{2\pi}d\phi'
\biggl[\sum_{M'=0}^{\infty}\Delta_{\bar{M}}\cos(\bar{M}\phi')\biggr]
\biggl[\sum_{M=0}^{\infty} F_{4M}(\Omega)\cos (4M\phi')
\exp\{-A|\phi-\phi'|\}\biggr]=
$$
$$
=\sum_{M,M'=0}^{\infty}F_{4M}\Delta_{\bar{M}}
\biggl[{A\over A^2+(4M+\bar{M})^2}\cos [(\bar{M}+4M)\phi]
+
{A\over A^2+(4M-\bar{M})^2}\cos [(\bar{M}-4M)\phi]\biggr]
\eqno(4)
$$

It is interesting to
consider first the isotropic part of the DOS ($M=0$) that reads    
$$
I_{\Delta}\approx {2A\over A^2+\bar{M}^2}\cos(\bar{M}\phi)
\eqno(5)
$$
which is perfectly symmetric to the damping contribution given
in equation (3) except that the order of the pairing harmonic $4M$ is 
replaced by the order of the gap harmonic $\bar{M}$. 
The contribution of the isotropic part of the interaction 
($M=0$) to an isotropic  
gap ($\bar{M}=0$) 
is therefore equal to its contribution to the damping, and this is the
conventional expectation in isotropic superconductivity.
However the contribution 
of the isotropic part of the interaction ($M=0$) to higher gap
harmonics ($\bar{M}\neq 0$) 
is smaller and this appears natural. Indeed we expect an isotropic
gap to be favored in the case of an isotropic electronic system.
We show in figure (2) the $\phi_0$ dependence of the 
contribution of the
$M=0$ DOS harmonic to the different gap harmonics.
The symmetry of equations 
(3) and (5)
gives a qualitative understanding of the electronic topology dependence 
of the damping effects. In fact
{\it the small 
damping contribution of the higher DOS harmonics ($M\neq0$) 
is equivalent to    
the smaller contribution of the isotropic
part of the DOS ($M=0$) to an anisotropic gap ($\bar{M}\neq0$)}.

We now consider the anisotropic part of the interaction and we analyze 
equation (4). If we have an isotropic gap ($\bar{M}=0$) the pairing  
contribution (eq. 4) is the same with the damping contribution (eq. 3).
The contribution of a higher DOS harmonic ($M\neq0$) to an isotropic gap
($\bar{M}=0$)
is {\it equal} to its contribution to the damping. Therefore, although the
anisotropic part of the interaction contributes less to the damping,
if the gap is isotropic it also contributes less to the
pairing and therefore we do not expect an enhancement of $T_c$ in that case.

The situation is totally different when the gap
is anisotropic and has higher order harmonics 
($\bar{M}\neq0$). The dominant term in equation
(4) is the second one an can be significant                         
when in the denominator $4M-\bar{M}=0$. Comparing equations (3) and (4)
it is easy to see that the contribution of a $M\neq0$ DOS harmonic 
to the pairing is higher to its contribution to the damping when 
$4M-\bar{M}=0$. We show in figure 3 the $\phi_0$ dependence of the
pairing amplitude of the $M=1$ DOS harmonic for different gap harmonics:
$\bar{M}=0$ (isotropic), $\bar{M}=2$ ($d_{x^2-y^2}$ gap) and $\bar{M}=4$
(anisotropic s component). The contribution to the $\bar{M}=0$ gap equals the 
damping contribution of this DOS harmonic. 
However the contribution of a given DOS $M\neq0$ harmonic
to the damping is not proportional to its contribution 
to $\bar{M}\neq0$ pairing.
In fact the contribution
of the $M=1$ DOS harmonic to the d-wave ($\bar{M}=2$) and 
anisotropic s-wave 
gap ($\bar{M}=4$) 
is {\it higher} than its contribution to the damping (compare figs. 1 and 3). 
This of course enhances significantly the $T_c$ contribution of this 
DOS
harmonic in the case of anisotropic gap.
The effect is amplified          
when we consider higher order harmonics of the DOS anisotropies
(higher values of M).
We also remark that the higher is the DOS harmonic the smaller is the 
characteristic angle $\phi_0$ from which the pairing contribution
becomes larger than the damping contribution of the interaction.

Of course all the previous contributions are weighted by the 
Fourier coefficients $F_{4M}$ and $\Delta_{\bar{M}}$.
The gap Fourier coefficients $\Delta_{\bar{M}}$ 
are obtained self consistently 
by the gap equation and therefore they also reflect the $F_{4M}$
coefficients and the
previously discussed pairing-damping contributions. 
The $F_{4M}$ coefficients which describe the
anisotropies of the electronic DOS are therefore the relevant
material parameter for our discussion.
To our approach the optimal situation for high-$T_c$ is 
the situation in which the weight of the higher order Fourier 
components $F_{4M}$ 
is the larger. This can be obtained for example when 
the van hove singularity 
is close to the Fermi level and the electronic density of states 
variates sharply. 

Our arguments for the enhancement
of $T_c$ are purely topological,
and provide therefore a new perspective to the van Hove singularity
approach to high-$T_c$.
Up to now,
the studies of the effect of the van Hove singularity in  
the self consistent Eliashberg 
framework have been done considering
a totally isotropic system and reported the impossibility of this
mechanism to produce a sharp enhancement of $T_c$ contrary to previous
claims based on simple BCS \cite{Radtke}.
In the light of our analysis, 
the effect on $T_c$ of the van Hove 
singularity will be amplified compared
to that predicted in an isotropic situation, since the effective 
enhancement of the coupling 
(enhancement of the $F_{4M\neq0}$) acts more on the
pairing than on the damping. 

We can also obtain within our picture some insight on the essential 
differences between the
n-doped and p-doped cuprates.
The ARPES obtained quasiparticle peak near $E_F$ 
in the normal state of
$Nd_{2-x}Ce_xCuO_{4-\delta}$ \cite{King} (which is an n-type superconductor)
is much broader than the corresponding peak in $YBa_2Cu_3O_7$ and
$BiSr_2CaCu_2O_8$ \cite{Olson}. In the case of $NCCO$ the van Hove
singularity lies far below the Fermi level (at $\approx 300meV$ below $E_F$)
and
the anisotropic part of the DOS                           
is very small. In that case the lowest order DOS harmonics are 
dominant and the
whole electron-phonon coupling contributes to the damping and therefore
the width of the quasiparticle peak (which reflects the damping effects)
is rather large. In the case of the
p-doped cuprates instead, the van Hove singularity is close to the
Fermi level and we have significant high order harmonics in the DOS 
that contribute little to the width of the quasiparticle peak
in the normal state since they are not effective for the damping.
On the other hand the critical temperatures are much higher in the
p-doped materials because the high DOS harmonics can have
a significant contribution to an anisotropic s-wave or d-wave gap without
contributing to the damping. This $T_c$ differences are present 
in our scheme even if the overall coupling
between the two types of materials is of the same order.

Notice that within our analysis the anisotropic s-wave gap
($\bar{M}=4M'$) is favored with respect to the anisotropic d-wave
gap ($\bar{M}=4M'+2$) because of the tetragonal DOS symmetry.
This is also the result of numerical simulations when the Coulomb
pseudopotential is neglected \cite{pr9507052}. 
The situation can be inverted when the Coulomb pseudopotential 
reaches a critical value which depends on its characteristic momentum
variations \cite{pr9511139}. 
We understand however clearly that the anisotropic gap may result
from anisotropic parts of the DOS (the isotropic DOS contributes very few
to an anisotropic gap as it is shown in fig. 2) and therefore we can also  
understand why in the n-doped cuprates the gap seems isotropic
while in the p-doped cuprates with the large $F_{4M\neq0}$ components
the gap is very anisotropic and probably in some of them is of anisotropic
d-wave type \cite{pr9511139}. Notice finally that other situations that may
lead to sharp variations in $k$-space of the electronic DOS on the
Fermi level like the presence of CDW or nesting, could also be within
our analysis favorable for high-$T_c$.

\newpage

{\Large \bf Figure Captions}
\vskip 1.5cm

{\bf Figure 1:}
The contribution to the damping of different DOS harmonics as 
a function of the characteristic range of the scattering $\phi_0$.
Full line corresponds to $M=0$, the long-dashed line to $M=1$ and
the short-dashed line to $M=2$
\vskip 1.0cm

{\bf figure 2:}
The contribution to the pairing of the $M=0$ DOS harmonic 
as a function of $\phi_0$. Full line corresponds to $\bar{M}=0$,
long dashed to $\bar{M}=2$ and short dashed to $\bar{M}=4$.
\vskip 1.0cm

{\bf figure 3:} Same as in figure 2 but for the $M=1$ DOS harmonic.
 

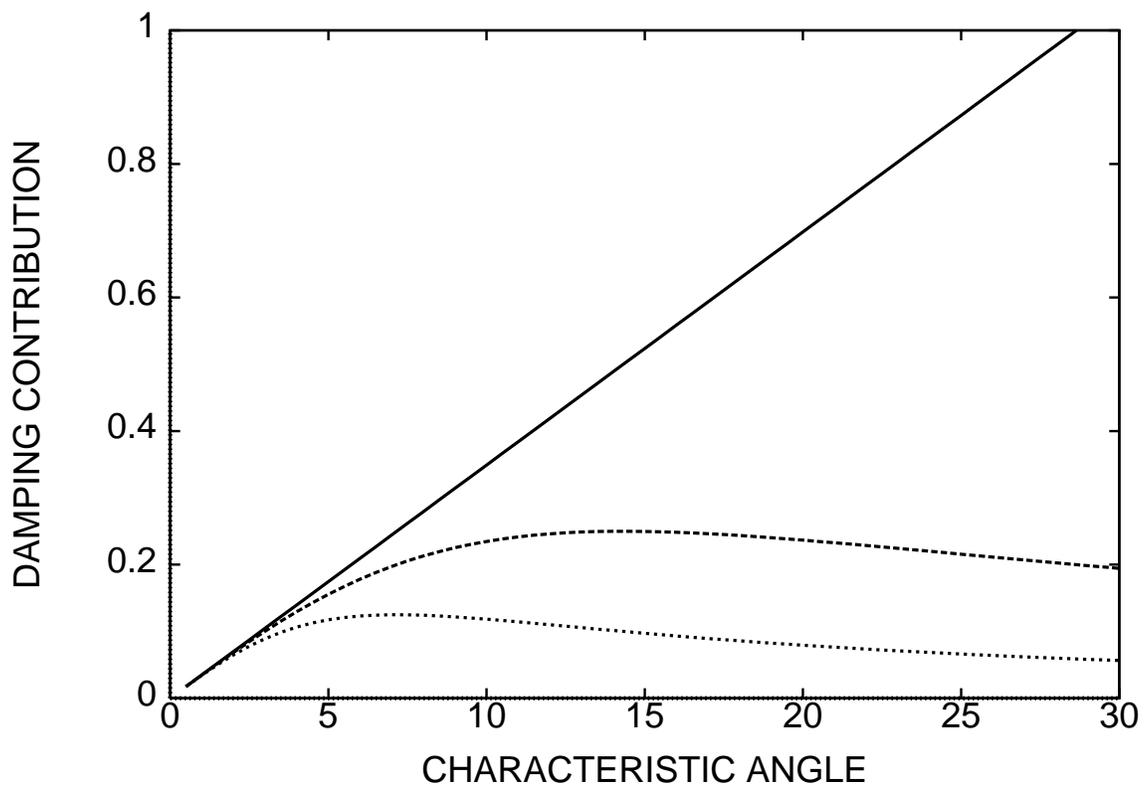
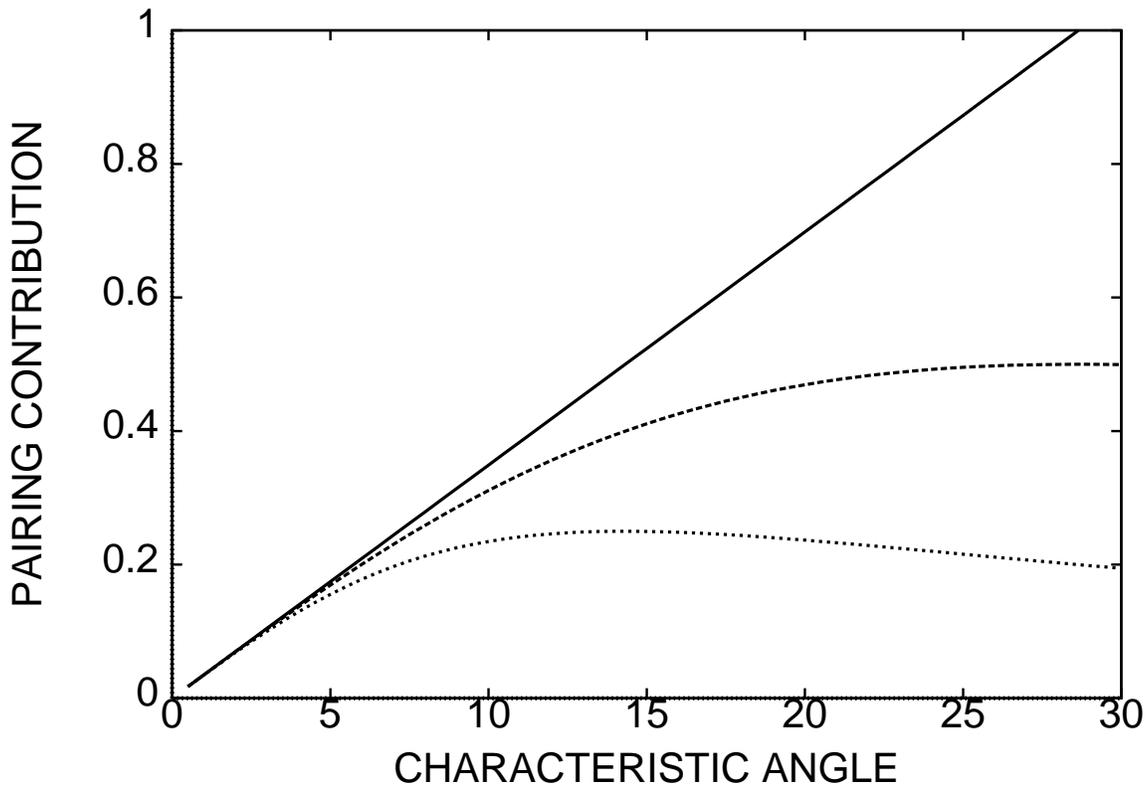
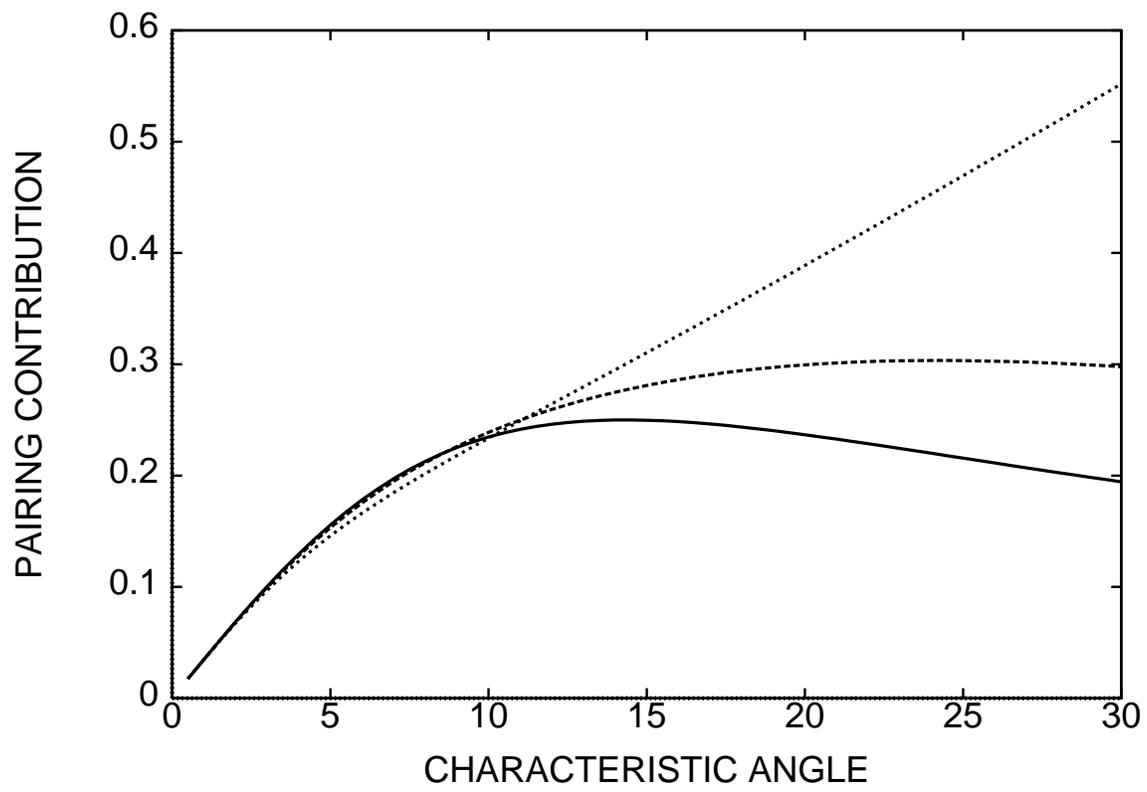
\end{document}